\documentclass[aps,prb,twocolumn,showpacs,showkeys,floatfix]{revtex4}

\usepackage{graphicx,color}
\usepackage{natbib}
\usepackage[utf8]{inputenc}
\usepackage{float}
\graphicspath{{figs/}}
\usepackage{amsmath}
\usepackage[colorlinks,linkcolor=blue,anchorcolor=blue,citecolor=blue]{hyperref}

\begin{document}

\title{Edge mode based graphene nanomechanical resonators for high-sensitivity mass sensor}

\author{Guang-Rong Han}
    \affiliation{Shanghai Institute of Applied Mathematics and Mechanics, Shanghai Key Laboratory of Mechanics in Energy Engineering, Shanghai University, Shanghai 200072, People's Republic of China}

\author{Jin-Wu Jiang}
    \altaffiliation{Corresponding author: jwjiang5918@hotmail.com}
    \affiliation{Shanghai Institute of Applied Mathematics and Mechanics, Shanghai Key Laboratory of Mechanics in Energy Engineering, Shanghai University, Shanghai 200072, People's Republic of China}

\date{\today}
\begin{abstract}

We perform both molecular dynamics simulations and theoretical analysis to study the sensitivity of the graphene nanomechanical resonator based mass sensors, which are actuated following the global extended mode or the localized edge mode. We find that the mass detection sensitivity corresponding to the edge mode is about three times higher than that corresponding to the extended mode. Our analytic derivations reveal that the enhancement of the sensitivity originates in the reduction of the effective mass for the edge mode due to its localizing feature.

\end{abstract}

\maketitle
\pagebreak


The nanomechanical resonator (NMR) has been proposed to be good mass sensors that are able to weigh cells, biomolecules, and gas molecules,\cite{Naik2009towards,Jensen2008atomic,Li2010nanolaeeters,Burg2007weighing} as the adhesion of external molecules lead to the shift of the resonant frequency of the resonator.\cite{WeipangAPL2006,Chaste2012nanomechanical,Hanay2015Inertial} Typically, the adhesion of masses will lead to the red-shift of the resonant frequency.\cite{Natsuki2013vibration,Berinskii2015differential} To achieve high-sensitivity mass sensor, resonators based on ultra light materials with high resonant frequency are most favorable.\cite{Chaste2012nanomechanical,Jensen2008atomic} Two dimensional materials attracted many interests because of its excellent mechanical properties (high Young's modulus, elasticity and breaking strength) and low mass. For example, \citet{Lee2013High} fabricated high frequency nanomechanical resonators based on $\rm{MoS_2}$ with thickness about 6~{nm} (a stack of 9 monolayers), which exhibit frequency up to 60~{MHz}. \citet{Castellanos2013Single} fabricated single-layer $\rm MoS_2$ mechanical resonators and the resonance frequencies are around 10~{MHz} to 30~{MHz}. \citet{Morell2016High} investigated the monolayer transition metal dichlcogenides (TMD) resonators at cryogenic temperatures, and find that the quality factor increases with decreasing temperature. Compared to these two-dimentional materials, graphene has larger Young's modulus, elasticity and breaking strength, and extremely low mass, which is of great important for mass sensing. In 2006, \citet{Bunch2007Electromechanical} fabricated nanomechanical resonators use single and multilayer graphenee sheets and find that the resonator has an ultimate limit on the force sensitivity of 0.9~$\rm {fN/Hz^{1/2}}$. Graphene resonantor's frequency can be tuned to the order of THz by decreasing its size to nanometer.\cite{Katsnelson2006Graphenetwodimen,LeeChanggu2008Science,NovoselovNature2012, AkinwandeExtreme2017} Furthermore, graphene has a larger surface area for the adsorption of molecules. As a result, the graphene nanomechanical resonator shall be a promising mass sensor. It was demonstrated experimentally that carbon nanotube resonator is able to detect the mass variation on the order of one gold atom.\cite{Jensen2008atomic,Lassagne2008Ultrasensitive,Chiu2008Atomic}

The resonant oscillation morphology of the NMR also plays an important role in the sensitivity of the NMR based mass sensor. The global extended mode and the localized mode are two typical modes that can be used to actuate the resonant oscillation of the NMR. In the global extended mode, all atoms are involved in the vibration, while only a group of atoms are involved in the vibration of the localized mode.\cite{Born1954dynamical} The edge mode is a particular type of localized mode, where only atoms near the edge are involved in the vibration. Experiments have illustrated that ultra-high sensitivity can be achieved in the mass detection by using the eigen vector of the localized edge mode in the silicon beam.\cite{Spletzer2006ultrasensitive,Spletzer2008highly,Pierre1987Localization,Pierre1986Localized,Chen1992On} However, most existing works on the graphene NMRs (GNMRs) are based on the global extended mode, but the GNMRs have not been actuated with the localized edge mode. In particular, it is still unclear that whether the edge mode based GNMRs will have higher sensitivity or not as compared with the global extended mode.

In this work, we perform molecular dynamics simulations to comparatively investigate the sensitivity of the mass detection using the GNMRs, which are actuated with the global extended mode or the localized edge mode. The mass sensitivity corresponding to the edge mode based resonator is about three times higher than the extended mode based resonator. The enhancement of the sensitivity can be attributed to the reduction of the effective mass in the localized edge mode as compared with the global extended mode. We derive analytical formula for the effective mass of the extended mode and the edge mode, which are in excellent agreement with numerical results.


The basic principle for sensing mass with nanomechanical resonators is the mass dependence for the resonant frequency. The resonant frequency will exhibit a red-shift when an adsorbate is adhered to the resonator. In general, the magnitude of the adsorbate induced frequency shift depends on the vibrational morphology of the resonator and the location of the adsorbate. For two-dimensional graphene, the frequency of the resonant oscillation is related to the mass of the system as follows,
\begin{equation}
f=R\frac{1}{\sqrt{m}}\label{1},
\end{equation}
where the parameter $R$ reflects material's properties of the resonator such as the bending rigidity and dimension aspect ratio. Some simple algebra gives the frequency shift ($\Delta f$) induced by the mass variation ($\Delta m$),
\begin{equation}
\Delta f=-\frac{f_{0}}{2m_{0}}\Delta m,
\label{deltaf}
\end{equation}
where $f_0$ and $m_0$ is the frequency and mass of the pure GNMR.

According to Eq.~(\ref{deltaf}), the coefficient $f_0/2m_0$ can be treated as the mass sensitivity. It is desirable to use a material of low density but with high resonant frequency $f_0$, so that high mass sensitivity can be achieved. The mass $m_0$ in Eq.~(\ref{deltaf}) actually represents the effective mass of the resonant vibration mode in the resonator. In previous works, the resonant mode excited in the graphene nanomechanical resonator has an extended vibrational morphology, i.e., all atoms are involved in the vibration, so the effective mass $m_0$ is indeed the total mass of graphene. However, in contrast to the extended mode, for localized modes, only a small portion of atoms are involved in the vibration, so the effective mass $m_0$ in Eq.~(\ref{deltaf}) represents the total mass of these vibrating atoms. In particular, the effective mass for the localized edge mode is much smaller than that of the extended mode. The edge mode in graphene is one particular localized mode, with atoms in the edge region involved in the vibration, so the effective mass for the edge mode is much smaller than the extended mode. As a result, the mass sensitivity ($f_0/2m_0$) for the graphene resonator oscillating with the edge mode's morphology shall be much higher than that of the extended mode. We will thus investigate possible enhancements for the mass sensitivity by exciting the edge mode in GNMRs.


\begin{figure}[tbp]
 \centering
 \scalebox{1}[1]{\includegraphics[width=8cm]{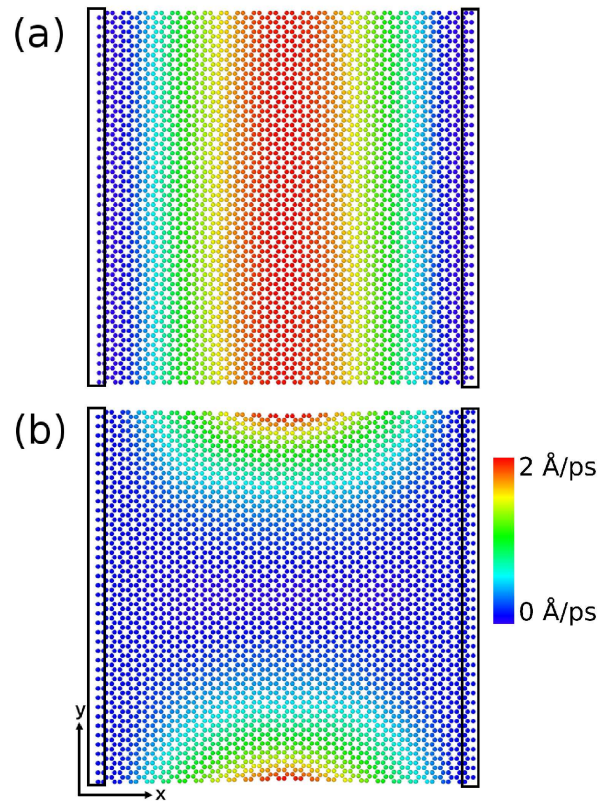}}
 \caption{GNMRs are actuated by (a) the extended mode with velocity distribution $v_z=v_0\sin (x/l_x)$ and (b) localized edge mode with velocity distribution $v_z=v_0\sin (x/l_x)e^{-y/l_c}$. The left and right ends are fixed during simulation.  Color bar represents the atomic velocity in the z-direction.}
\label{structure}
\end{figure}

Fig.~\ref{structure} shows the structure of the graphene simulated in the present work. The dimension is $l_x\times l_y=100\times 100$~{\AA}. The left and right ends (of 5.0~{\AA} at each end) are fixed during the simulation, while periodic or free boundary conditions are applied in the vertical y-direction. The adsorbate is simulated by the copper atoms. The interaction for graphene is described by the adaptive intermolecular reactive empirical bond order (AIREBO)\cite{Stuartpotential}, and the interaction among copper atoms is described by the embedded-atom method (EAM) potential.\cite{DawEAM1,DawEmbedded} Copper and carbon atoms interact through the Lennard-Jones potential,
\begin{equation}
V_{LJ}=4\epsilon\left[\left(\frac{\sigma}{r}\right)^{12}-(\frac{\sigma}{r})^{6}\right],
\label{eq_lj}
\end{equation}
with $\epsilon=25.78$~{meV} and $\sigma=3.08$~{\AA}.\cite{Wolf1989structurally} The standard Newton equations of motion are integrated in time using the velocity Verlet algorithm with a time step of 1~{fs}. Molecular dynamics simulations are performed using the publicly available simulation code LAMMPS\cite{PlimptonSJ,lammps}. The OVITO package was used for visualization\cite{Stukowski2009ovito}.

We will comparatively study the resonant oscillation based on the extended mode and the localized edge mode. There are three major steps for the simulation of the GNMR that oscillates following the extended mode, where periodic boundary condition is applied in the vertical y-direction in Fig.~\ref{structure}~(a). First, the system is thermalized to a constant pressure (0.0~Pa) and temperature (4.2~K) within the NPT (constant particle number, constant pressure, and constant temperature) ensemble for 1000~{ps} by the Nos\'e-Hoover approach.\cite{Nos1984A,Hoover1985canonical} Second, the resonant oscillation is actuated by adding a velocity distribution $v_z=v_0\sin(x/L_x)$ to the system, which follows the form of the extended mode. The amplitude $v_0$ is referred as the actuation velocity. The velocity distribution is illustrated in Fig.~\ref{structure}~(a). Third, the system is allowed to oscillate within the NVE (i.e., the particles number N, the volume V, and the energy E of the system are constant) ensemble. The dissipation of the resonant oscillation energy from the third step is used to analyze the mechanical oscillation of the system.

Slightly different simulation steps are applied for the simulation of the GNMR following the morphology of the localized edge mode. Free boundary condition is applied in the y-direction in Fig.~\ref{structure}~(b), so there are edge modes localized at the two free y-edges.\cite{Jiang2012edge} The system is thermalized within the NPT ensemble for 1000~{ps}, and graphene is stretched by 0.8\% so that the warped edges\cite{Fasolino2007Intrinsic,ShenoyPRL2008,Rafiee2010Fracture} can be flattened.\cite{JiangJAP2016} The resonant oscillation is then actuated by adding a velocity distribution $v_z=v_0\sin(x/L_x)e^{-y/l_c}$ to the system, where $l_c$ is a penetration depth and $y$ is with reference to the edge.\cite{Jiang2012edge} This velocity distribution is illustrated in Fig.~\ref{structure}~(b), which follows the vibrational morphology of the localized edge mode.

\begin{figure}[htpb]
 \centering
 \scalebox{1}[1]{\includegraphics[width=8cm]{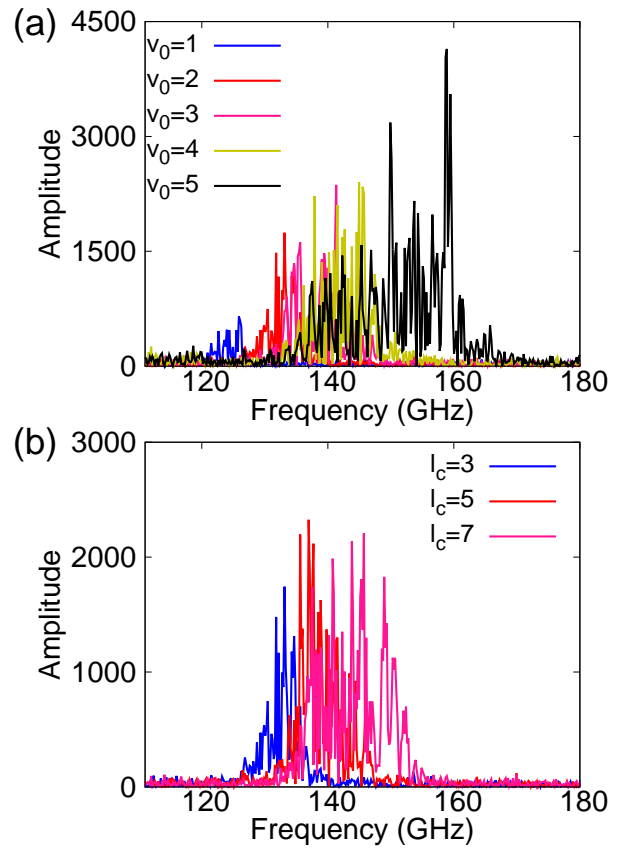}}
 \caption{The effects of actuation velocity and penetration depth on the resonant oscillation of GNMRs following the morphology of edge mode. The y-axis shows the Fourier transformation of the time history of the total kinetic energy. (a) The resonant oscillation is actuated with $l_c=3.0$~{\AA} and different $v_0$. (b) The resonant oscillation is actuated with $v_0=2.0$~{\AA/ps} and $l_c=$ 3.0, 5.0, and 7.0~{\AA}.}
\label{velocity}
\end{figure}

In the actuation of the edge mode, the actuation velocity $v_0$ and the penetration depth $l_c$ are two key parameters. We will first examine their effects on the resonant frequency. To study the effect of the actuation velocity, the resonant oscillation is actuated following the edge mode with penetration depth $l_c=3$~{\AA}, while the actuation velocity $v_0=$ 1.0, 2.0, 3.0, 4.0, and 5.0~{\AA/ps}. Results of this simulation set are shown in Fig.~\ref{velocity}~(a). The data shown in the figure is the Fourier transform of the time history of the kinetic energy. The central position of the peak reflects the frequency (actually twice of the frequency) of the resonant oscillation. There is an obvious blue-shift of the frequency with the increase of the actuation velocity, as a result of stronger nonlinear effect caused by larger actuation velocity. It seems that larger actuation velocity is favorable, as higher frequency leads to higher sensitivity. However, with the increase of actuation velocity, the Fourier peak will be broadened considerably, leading to larger error in extracting the frequency value. We thus use a moderate value of $v_0=2.0$~{\AA/ps} for the following simulations.

To study the effect of the penetration depth on the simulation, we actuate the resonant oscillation following the edge mode with actuation velocity $v_0=2.0$~{\AA/ps}, and the penetration depth $l_c=$ 3.0, 5.0 and 7.0~{\AA}. Fig.~\ref{velocity}~(b) shows that the width of the Fourier peak will be broadened with the increase of the penetration depth, which is not favorable. We thus use $l_c=3.0$~{\AA} in the following simulations.

\begin{figure}[tbp]
 \centering
 \scalebox{1}[1]{\includegraphics[width=8cm]{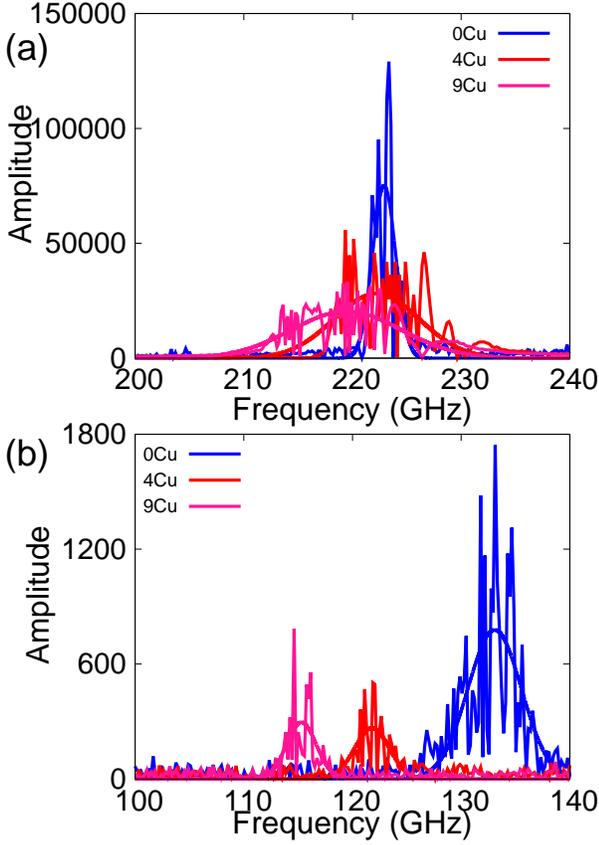}}
 \caption{Adsorbate induced red-shift for the resonant frequency in GNMRs following the morphology of (a) the extended mode and (b) the localized edge mode. Lines are Gaussian fitting results.}
\label{comparison}
\end{figure}

\begin{figure}[tbp]
 \centering
 \scalebox{1}[1]{\includegraphics[width=8cm]{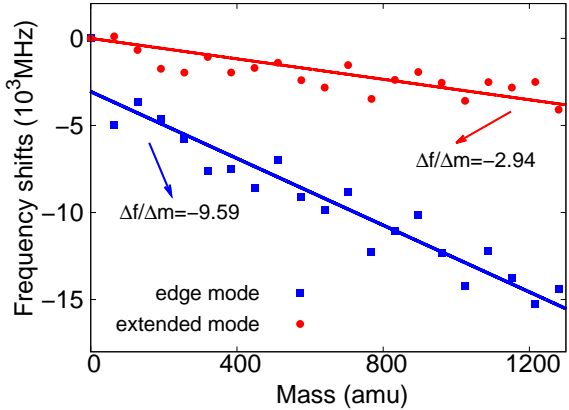}}
 \caption{The frequency shift versus the adsorbate mass. }
\label{frequencymass}
\end{figure}

We now compare the mass sensitivity of the GNMRs working based on the extended mode or the localized edge mode. As shown in Fig.~\ref{comparison}, the resonant frequency will be reduced more by adsorbing more copper atoms on graphene. The frequency shift in Fig.~\ref{comparison}~(b) for the GNMR based on the edge mode is obviously larger than Fig.~\ref{comparison}~(a) for the GNMR based on the extended mode. The exact value of the frequency is extracted by doing a Gaussian fitting to the Fourier peak. The frequency shift is illustrated in Fig.~\ref{frequencymass} for the resonant oscillation following the extended mode or the edge mode. A linear fitting yields the mass sensitivity of 2.94~{MHz/amu} and 9.59~{MHz/amu} for the GNMRs based on the extended mode and the edge mode, respectively. Our molecular dynamics simulations show that the mass sensor based on the edge mode is about three times more sensitive than the mass sensor based on the extended mode.

We will now provide some analytic explanations for the enhancement of the mass sensitivity by actuating the resonator using the edge mode. According to Eq.~(\ref{deltaf}), the mass sensitivity is $\Delta f/\Delta m=f_0/2m_0$, where the mass $\Delta m$ is the variation of the total mass. In other words, the mass variation $\Delta m$ is uniformly distributed over the whole GNMR. However, in the above numerical simulations and also in practical applications, adsorbates are adhered to some local regions of the GNMR, so some structural corrections are needed for the expression of the mass sensitivity. As a result, the mass sensitivity shall be $\Delta f/\Delta m=(f_0/2m_0)S$, where $S=1-cos(2 \pi x/l_x)$, with $x$ as the position of the adsorbate, is the structural correction factor proposed by \citet{Dai2009Nanomechanical}. The physical meaning of this structural correction factor is that an adsorbate at $x=l_x/2$ (with maximum oscillation amplitude) causes the strongest influence on the frequency with $S=2.0$; while an adsorbate near two fixed ends ($x=0$ or $l_x$) induces neglectable reduction on the frequency with $S=0.0$. In this work, the adsorbs locate at $x=l_x/2$, so the structural correction factor is $S=2.0$. 

For the extended mode, the effective mass $m_0=0.9 m_g$ with $m_g$ as the total mass of graphene. A factor of 0.9 is to exclude the two fixed regions in the x-direction. The frequency $f_0=112.46$~{GHz}. After some simple algebra, we obtain the mass sensitivity of the extended mode $(f_0/2m_0)S=2.86$~{MHz/amu}. This analytic result is in good agreement with the numerical simulation result of 2.94~{MHz/amu}.

\begin{figure}[tbp]
 \centering
 \scalebox{1}[1]{\includegraphics[width=8cm]{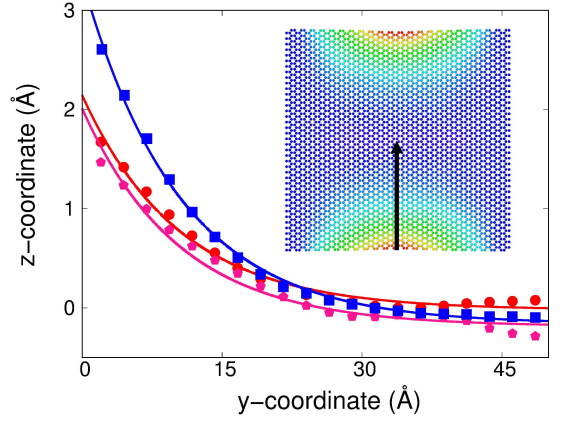}}
 \caption{The coordinates (y, z) for atoms at $x=l_x/2$ (as indicated by the arrow in the inset). Three sets of data are the structure at three different simulation time. These three data sets are fitted to $z=a e^{(-\frac{y}{l'_c})}+b$ with the same value of $l'_c=10.44$~{\AA}. Here y is with reference to the lower boundary of the inset.}
\label{position}
\end{figure}

For the edge mode, the effective mass $m_0$ is computed as follows
\begin{equation}
m_0=0.9\times m_g \frac{2}{l_y} \int_{0}^{\frac{l_y}{2}}e^{-y/l'_c}dy,
\label{eq_m0_edge}
\end{equation}
where the integration over y is performed in half of the graphene plane with $y\in[0, l_y/2]$. The exponential function is the shape of the graphene plate during the resonant oscillation following the morphology of the edge mode. The critical length $l'_c$ is obtained from the snapshots from the molecular dynamics simulation. Fig.~\ref{position} displays the coordinates (y, z) for atoms at the middle vertical line with $x=l_x/2$. Positions of these atoms are varying during the resonant oscillation of the system. We have shown the positions at three different simulation times, which can be well fitted to the exponential function with the same critical length $l'_c=10.44$~{\AA}. With this critical length, we can thus extract that $m_0=0.9 m_g\times 0.21$ for the localized edge mode from Eq.~(\ref{eq_m0_edge}). The frequency for the pure GNMR actuated by the edge mode is $f_0=66.74$~{GHz}. As a result, the mass sensitivity for the edge mode based GNMR is $(f_0/2m_0)S=8.12$~{MHz/amu}, which agrees quite well with the numerical result of 9.59~{MHz/amu}. This analytic value for the mass sensitivity is about three times larger than the analytic value of 2.86~{MHz/amu} for the extended mode based GNMR.

We have thus demonstrated both numerically and analytically that the mass sensitivity of the edge mode based GNMR is about three times higher than the extended mode based GNMR. According to Eq.~(\ref{deltaf}), high frequency and low effective mass are required for a mass sensor of high sensitivity. We find that the frequency (66.74~GHz) of the resonator based on the localized edge mode is slightly lower than the frequency (112.46~GHz) of the resonator based on the extended mode. However, from the analytic derivation, it is clear that this sensitivity enhancement is achieved through the localization property of the edge mode as illustrated by Eq.~(\ref{eq_m0_edge}). More specifically, the effective mass for the edge mode ($0.9 m_g\times 0.21$) is only one fifth of the effective mass for the extended mode ($0.9 m_g$), which directly leads to the great enhancement of the mass sensitivity according to Eq.~(\ref{deltaf}).

It should be noted that temperature plays an important role for the nanomechanical resonator. The energy of the resonator dissipates faster at higher temperature. It is because the nonlinear phonon-phonon scattering is stronger at higher temperature, which leads to the dissipation of the resonant energy. There is a common diffusion issue for the mass detection at high temperature; i.e., adsorbates will diffuse around the surface of the graphene resonator at high temperatures, as the interaction between the adsorbate and graphene is weak.


In conclusion, we have investigated the ability of detecting masses on the atomic level for the GNMR, which works based on the extended mode or the localized edge mode. As compared with the extended mode, the mass sensitivity of the GNMR can be upgraded by a factor of three by using the localized edge mode as the oscillation morphology. We derive the analytic results for the mass sensitivity, which discloses the underlying mechanism for the high mass sensitivity to be the much smaller effective mass of the localized edge mode as compared with the extended mode.

\textbf{Acknowledgment} The work is supported by the Recruitment Program of Global Youth Experts of China, the National Natural Science Foundation of China (NSFC) under Grant No. 11504225, the start-up funding from Shanghai University, and the Innovation Program of Shanghai Municipal Education Commission under Grant No. 2017-01-07-00-09-E00019.

\bibliographystyle{apsrev}

\end{document}